

%
\catcode`@=11 
\def\Number#1{\number #1}
\newcount\chapternumber      \chapternumber=0
\newcount\equanumber         \equanumber=0
\let\chapterlabel=0
\newtoks\chapterstyle        \chapterstyle={\Number}
\def\makel@bel{\xdef\chapterlabel{%
\the\chapterstyle{\the\chapternumber}.}}
\def\setchapternumber#1{\chapternumber=#1 \makel@bel}
\def\eqname#1{\relax \ifnum\equanumber<0
     \xdef#1{{\rm(\number-\equanumber)}}\global\advance\equanumber by -1
    \else \global\advance\equanumber by 1
      \xdef#1{{\rm(\chapterlabel \number\equanumber)}} \fi}
\def\eq{\eqno\eqname\?\?}
\def\eqn#1{\eqno\eqname{#1}#1}

\def\eqinsert#1{\noalign{\dimen@=\prevdepth \nointerlineskip
   \setbox0=\hbox to\displaywidth{\hfil #1}
   \vbox to 0pt{\vss\hbox{$\!\box0\!$}\kern-0.5\baselineskip}
   \prevdepth=\dimen@}}
\def\sequentialequations#1{\equanumber=-#1}
\def\nextequation#1{\global\equanumber=#1
   \ifnum\the\equanumber>0 \global\advance\equanumber by -1 \fi}
\catcode`@=12 

%
%

\def\VEV#1{\left\langle{#1}\right\rangle}              
   %
\def\ee#1{{\rm e}^{#1}}                                
\def\abs#1{\left\vert #1\right\vert}                   
\def\gbf#1{\setbox0=\hbox{$ {#1}$}		       
\kern-.025em\copy0\kern-\wd0                  	       %
\kern.05em\copy0\kern-\wd0                             %
\kern-.025em\raise.0433em\box0 }                       %
\def\lapp{\raise 1pt\hbox{\spose {$\scriptscriptstyle <$}}
\lower 2pt\hbox{$\scriptscriptstyle \sim$}}
\def\gapp{\raise 1pt\hbox{\spose {$\scriptscriptstyle >$}}
\lower 2pt\hbox{$\scriptscriptstyle \sim$}}

%
\def\dalembertian{\vbox{\hrule\hbox{\vrule\kern .7ex
	\vbox{\kern .7ex \kern .7ex}\kern .7ex\vrule}\hrule}}
\def\frac#1#2{ {{#1}\over{#2}} }

\def\inverse#1{ \frac{1}{#1} }
\def\subr#1{_{\rm #1}}

%
%
\def\cbf#1{\centerline{\bf #1}}                        

\def\etal{{\it et al.} }                               

\def\leaderfill{\leaders\hbox to .91em{                
\hss .\hss}\hfill}                                      %
\def\tabcon#1#2{\hbox to \hsize{#1 \leaderfill         
#2}}                                                   %
\def\undertext#1{\vtop{\hbox{#1}\kern 1pt \hrule}}

\def\todaysdate{\the\day\tie\monthword\tie\the\year}

%
%
\def\refskip#1{\gdef\r@fskip{#1}}
\def\qq{\par\r@fskip\noindent\hangindent .5in {\hangafter 1} }
\refskip{\smallskip}

%

%


\tolerance=1000
\sequentialequations{1}

\def\xir{\xi (r)}
\def\pk{P(k)}
\def\ct{C(\theta )}
\def\vv{\VEV{v^2_p(R)}}



\null \vskip -3ex


\cbf{HOW COSMIC BACKGROUND CORRELATIONS AT LARGE ANGLES}
\cbf{RELATE TO MASS AUTOCORRELATIONS IN SPACE}

\bigskip\bigskip
\centerline{GEORGE R. BLUMENTHAL and KATHRYN V. JOHNSTON}
\centerline{UCO/Lick Observatory}
\centerline{Board of Studies in Astronomy and Astrophysics}
\centerline{University of California, Santa Cruz}
\centerline{Santa Cruz, California 95064}

\vskip .3in
\cbf{ABSTRACT}

The Sachs-Wolfe effect is known to produce large angular scale
fluctuations in the Cosmic Microwave Background Radiation (CMBR) due
to gravitational potential fluctuations. We show how the angular
correlation function of the CMBR can be expressed explicitly in terms
of the mass autocorrelation function $\xi (r)$ in the Universe.  We derive
analytic expressions for the angular correlation function and its
multipole moments in terms of integrals over $\xi (r)$ or its
second moment, $J_3 (r)$,
which does not need to satisfy the sort of integral constraint that
$\xi (r)$ must. We derive similar expressions for bulk flow velocity
in terms of $\xi$ and $J_3$. One interesting result that emerges
directly from this analysis is that, for angles $\theta$, there is
a substantial contribution to the correlation function from a wide
range of distance $r$ and that the radial shape of this contribution
does not vary greatly with angle.

\noindent
{\bf Subject Headings:} Cosmic Microwave Background --- Cosmology:
Large Scale Structure of the Universe.

\vskip .3in
\line{{\bf 1~~ Introduction.} \hfil}

One sensible explanation for the currently clumpy state of the
Universe is that it arose from the gravitational amplification of an
initially nonuniform background mass density.  The anisotropy of the
Cosmic Microwave Background Radiation (CMBR) provides a signature of
these primordial density perturbations. In simple terms, variations in
the gravitational potential arising from density fluctuations at the
last scattering surface produce
variations in the redshift of photons across the sky.  This is the
Sachs-Wolfe effect (Sachs and Wolfe 1967, Wolfe 1968), which is the
dominant contributor to CMBR fluctuations on large angular scales.

The CMBR anisotropy is most easily quantified in terms of the angular
correlation function, $\ct = \langle \delta T_1 \delta T_2 \rangle / 
\langle T \rangle ^2$, of the temperature averaged over all directions in
the sky; here $\theta$ is the angle between the directions 1 and 2.
For potential fluctuations in a perturbed, flat, Friedmann universe,
Sachs and Wolfe (1967) related $\ct$ to the present power spectrum of
matter fluctuations, $\pk$, by following photon paths from the last
scattering surface to the observer.  Since the power spectrum
itself is just given by the Fourier transform of the mass
autocorrelation function $\xi (r)$, it is also possible to express
$\ct$ in terms of the autocorrelation function. This was done for some
special cases by Traschen and Eardley (1986).  

It is also well known
that there are several other effects which can and do contribute to
the CMBR anisotropy. These include Doppler shifts and thermalization
effects associated with the surface of last scattering. However, on
large angular scales, the dominant contributor to fluctuations is the
Sachs-Wolfe effect.

In discussing the angular correlations of the CMBR, it is traditional
to subtract out the contribution to $\ct$ arising from its monopole
and dipole moments (cf. Martinez-Gonzalez and Sanz 1989; Gorski 1991).
The monopole term is not physically interesting, while the dipole
contribution to the observed correlation of temperature is usually
assumed to be dominated by the peculiar motion of the observer with
respect to the local standard of rest. In that case the dipole
velocity is not itself directly related to potential fluctuations.
The recent positive detection by COBE of fluctuations in the CMBR
(Smoot \etal, 1992) has been reported with monopole, dipole and
quadrupole moments of $\ct$ removed to exclude these large
contributions to the anisotropy from sources other than the
Sachs-Wolfe effect. For example, the Milky Way itself is a strong
quadrupole source.

Another measure of large scale structure in the Universe which depends
explicitly on the power spectrum $\pk$ is the large scale bulk
motion in the Universe (Lynden-Bell \etal 1988, Gorski 1991, Dekel \etal
1990). A particularly useful quantity describing various models is the
mean squared peculiar velocity $\vv$, averaged over some scale $R$.

In this paper, we derive explicit expressions for the peculiar
velocity $\vv$ as well as the temperature
correlation function $\ct$, both with and without its first few
moments removed, as integrals over the mass autocorrelation function,
$\xir$ or its own integral $J_3(r)$.
We also derive similar expressions for each term in the multipole
expansion of $\ct$.
In Section 2 we derive the basic relations, while in Section 3, we
discuss how the various correlation functions interrelate.

\bigskip
\medskip
\line{{\bf 2 Fundamental Relations} \hfil}
\nobreak
\line{{\bf 2.1 Definitions} \hfil}
\nobreak

In any N-dimensional space, the correlation function is a useful
statistic of quantifying some of the relationships between values of a
function at different spatial positions. The function can be
continuous, as in the case of fluctuations in temperature or density,
or it can even be discrete.  A conceptually easy example of the latter
is that of the autocovariance function of galaxies in three
dimensions. Here it is a measure of the probability of finding a
neighbor in the data set at a given distance from any galaxy (see
Peebles, 1981, for example).

The correlation function of any continuous distribution is defined
mathematically in terms of an average over all of N-dimensional space. 
Hence, the {\it
angular correlation} of temperature fluctuations in the CMBR on the
scale $\theta$ is defined as 
$$ 
  C(\theta )= {\langle \delta T_1 \delta
   T_2 \rangle \over \langle T \rangle ^2}, \eqn\coftheta 
$$ 
and the {\it
mass autocorrelation} function for matter a distance $r$ apart is given by 
$$
  \xi (r)= {\langle \delta \rho_1 \delta \rho_2 \rangle \over \langle
  \rho \rangle ^2}.  \eqn\xiofrdef 
$$ 
The quantities $\delta T$ and
$\delta \rho$ are respectively the temperature and density excess over
the mean,  with subscripts $1$ and $2$ denoting position separated by an
angle $\theta$ or distance $r$. The angular brackets signify averaging
in two dimensions over the sky and over all three dimensional space.
It is assumed that the correlations are independent of direction.

The {\it power spectrum} $P(k)$ is a measure of the amplitude of density
perturbations in Fourier space for a given wave-number. It is defined
simply as $\abs{\delta_k}^2$, where $\delta_k$ is the Fourier transform of
the dimensionless density excess ${\delta \rho / \langle \rho
\rangle}$. As a result
of these definitions, it is obvious that 
$P(k)$ is the three dimensional Fourier
transform of $\xir$. 
Since the autocorrelation function is independent of direction, one can write
$$
     P(k)=\abs{\delta_k}^2 =4\pi \int_{0}^{\infty} dr\, r^2\, 
     \xi (r) \,{\sin kr \over kr}. \eqn\pofkdef
$$
The power spectrum \pofkdef\ often arises in expressions for observable
quantities which arise from density or potential 
fluctuations. Any such observable quantity, $Y$, usually
involves integrals in
Fourier space of the form
$$
       Y(x)=\int_{0}^{\infty} dk\, k^2\, P(k)\, F_Y(k,x).\eq
$$
The quantity $ F_Y(k,x)$ explicitly relates the observed quantity $Y$
(which may depend on some variable $x$) to the amplitude of waves
having wavenumber $k$.

We can relate the observed quantity $Y$ to the autocorrelation function by
substituting for $P(k)$ from equation \pofkdef\ and performing the
$k$-integral to yield
an expression involving an integral in real space. In this way we 
derive the {\it kernel}, $G_Y(r,x)$, for
mass fluctuations, peculiar velocities and temperature variations in
the CMBR, such that
$$
     Y(x)=\int_{0}^{\infty} dr\, r^2\, \xi (r) \, G_Y(r,x).\eqn\ykernel
$$
This kernel is  related directly to $F_Y$ by the expression
$$
     G_Y(r,x)={4\pi \over r} \int_{0}^{\infty} dk\, k\, \sin(kr)\, 
     F_Y(k,x),\eqn\gfrel
$$
or by the inverse expression
$$
     F_Y(k,x)={1 \over 2\, \pi^2\, k} \int_{0}^{\infty} dr\, r \,
     \sin(kr)\, G_Y(r,x).\eq
$$ 
Integration by parts of equation \ykernel\ leads to
$$  
     Y(x)=-\int_{0}^{\infty} dr \, J_3(r) \, {\partial G_Y \over
     \partial r}, \eqn\yjjj
$$
where 
$$
   J_3(r)=\int_{0}^{r}dr\;r^2\xi(r). \eq
$$ 
Since by definition, $\xi(r)$ must satisfy the integral constraint
$J_3(\infty)=0$, while this is simply a boundary condition on $J_3$,
one can gain 
some insight into the underlying physics by considering (as in
equation \yjjj) the derivative of the
kernel $\partial G_Y / \partial r$ rather than the kernel $G_Y$ itself.

\bigskip
\line{{\bf 2.2 Mass Fluctuations and Peculiar Velocities.} \hfil}
\nobreak

Consider a sphere having radius $R$. Then the mean squared mass excess
in a randomly placed sphere of this radius is given by
$$
   \VEV{\left( {\delta M \over M}(R) \right)^2} = 
   {1 \over 2 \pi ^2} \int_{0}^{\infty} dk\, k^2\, P(k)\, W^2(kR),
   \eqn\dmmr  
$$
where $W(kR)$ is the window function, which is given by a Fourier
transform of a three dimensional top hat sphere of radius $R$:
$$
   W(kR)= {3\over{4\pi R^3}} \int_{\rm sphere} d^3{\bf r}\, \ee{i {\bf
   k}\cdot {\bf r}} =
   {3 \over (kR)^3}[\sin kR -kR \cos kR] =
   3 {j_1(kR) \over kR}, \eqn\massth
$$
where $j_1$ is a spherical Bessel function.
In analogy with eq. \ykernel, we can 
rewrite equation \dmmr\ as 
$$
   \VEV{\left( {\delta M \over M}(R) \right)^2} =
   \int_0^\infty r^2 dr \xi (r) G_{\delta M} (r,R) , \eq
$$
where equation \gfrel\ allows us to explicitly evaluate $G_{\delta M}$ as
$$
    G_{\delta M} (r,R) = {3 \over 2R^3} 
    \cases {2-3x+x^3, &if $x < 1$ \cr   0, &otherwise, \cr}\eqn\gdeltam
$$
where  $x=r/2R$. Clearly, $G_{\delta M}$ goes to zero for $r>2R$ since
correlations larger than the diameter of the sphere are irrelevant.

As an alternative to the top hat expression (\massth), 
one can smooth the power spectrum by weighting 
the points with a Gaussian 
window function of the form
$$									
   W(kR_g) = \ee{-k^2 R\subr{g}^2 /2 } , \eqn\gaussdef
$$
where $R\subr{g}$ is the Gaussian smoothing length. 
For this Gaussian window function,
equation \gfrel\ allows us to obtain the kernel
$$
      G_{\delta M} (r,R) = \inverse{2\sqrt{\pi} R\subr{g}^3}
      \ee{-  r^2/4R\subr{g}^2 } .\eqn\gdeltamg
$$
In a sense, \gdeltamg\ is obvious since it is again the window function
in real space. Some
papers relate the Gaussian smoothing length to the top hat sphere
radius. For example, Blumenthal \etal (1992) require the Gaussian
window to have the same spatial volume as the top hat sphere:
$$
    R\subr{g} = \left( \frac{2}{9\pi} \right)^{1/6} R, \eqn\rgr
$$
while Lecar (1993) points out that for $R_g \approx .47 R$, the
Gaussian window function gives similar integrals to the top hat for a
wide range of power spectra.

The kernels $G_{\delta M}$ for both the top hat window function and
for the  Gaussian window with $R\subr{g}$ given by \rgr\ are plotted
in Figure 1a.  Note that  the expression for $\delta M/ M$ follows
directly from the definition of $\xi (r)$ rather than from an evolutionary
calculation. Thus the kernel for the top-hat window is zero for
distances larger
than the distance across the top-hat ($2R$) since by definition, mass
distributions are not sampled on these scales. 
The kernel for the
Gaussian extends to cover all space since the weighted average is
taken on all scales.
In Figure 1b we have plotted the derivative of the kernel, which is 
used with the $J_3$ integral \yjjj. 

In a similar fashion, one can calculate, to first order in perturbation
theory, the 
rms {\it peculiar velocity} averaged over randomly placed 
regions of size $R$:
$$
   \vv =  {H_o^2 \, {\cal F}^2 (\Omega)
   \over 2 \pi ^2} \int_{0}^{\infty} dk\, k^2\, P(k)\, 
   {W^2(kR) \over k^2},\eqn\pecvel
$$
where ${\cal F}(\Omega)$ is well approximated by $\Omega^{0.6}$ for density
parameter $0<\Omega <1$ (Peebles 1976) and is approximately
$\Omega^{4/7} $ for $\Omega\approx 1$ (Lightman and Schechter, 1991),
and $\Omega$ is the
dimensionless density parameter. Actually, ${\cal F}$ represents the
logarithmic rate of change of the amplitude of fluctuations with scale
factor. 
The expression \pecvel\ is valid when the cosmological constant in
zero, and ${\cal F}$ must be modified in the case of a nonzero
cosmological constant. $H_o$ is the current 
value of Hubble's parameter. For a top-hat window, \gfrel\ yields the
peculiar velocity kernel:
$$
      G\subr{v} (r,R) =  {H_o^2 {\cal F}^2 (\Omega ) \over R}\cases { 
      ({6 \over 5}
      -2x^2+{3\over2}x^3-{1\over 5}x^5), &if $x < 1$ \cr 
      {1 \over 2 x}, &otherwise.\cr}
      \eqn\gsigma
$$
The comparable expression using the 
Gaussian window-function \gaussdef\ is given by
$$
      G\subr{v} (r,R) = {H_o^2 {\cal F}^2 ( \Omega ) \over r}\;
      {{\rm erf} \left( r/2R\subr{g} \right)},\eqn\gsigmag
$$
where erf is the error function.

The velocity kernels $G\subr{v}$ (and their derivatives) for both the
top-hat window and the Gaussian window with $R\subr{g}$ given by \rgr\
are also plotted in Figure 1.
Here, the difference between top hat and Gaussian is more pronounced
in the kernel derivative (Figure 1b) than in the kernel itself. This
is a reflection of the fact that the autocorrelation function is in
some sense a
measure of the density distribution while $J_3$ measures the mass
distribution and it is on this that the peculiar velocity is more
heavily dependent. 
The kernel falls as $r^{-1}$ on large scales ($r\gg R$). This
behavior is obvious since the  large scale contribution goes as
$\VEV{v^2_p} \propto M(r)/r \propto \rho  J_3(r) /r \propto \rho \xi(r)
r^2 dr/r$, and large regions can have a significant effect on the
velocity, as can be seen already from equation \pecvel. It is also
similar to the 
expression for mean squared velocity in Peebles 1981 (equation 14.10),
$$
   \langle v^2_p \rangle = H_0^2 {\cal F} ^2(\Omega) \int_{0}^{\infty} 
   dr\, r\, \xi (r),
   \eq
$$
which explains the limiting behavior of $G\subr{v} \rightarrow H_0^2 {\cal
F}^2(\Omega)/2Rx$.

Kashlinsky (1992) used a similar method to relate both the peculiar velocity
correlation function ($\nu (r)=\langle {{\bf v}_p}_1 . 
{{\bf v}_p}_2 \rangle$) to
$\xi (r)$ and the {\it rms} peculiar velocity to the projected angular
correlation function of matter, $w(\theta)$, found in the APM survey
(Maddox \etal,1990).

\bigskip

\line{{\bf 2.3 Angular Correlation Function.} \hfil}
\nobreak

The Sachs-Wolfe result for the CMBR angular correlation function is given by
$$
   \eqalign{C(\theta ) 
   &={H_o^4 {\cal F} (\Omega)
      \over 8\pi ^2 c^4} \int_0^{\infty} {dk \over k^2}\, P(k)\, 
      {\sin ky\over ky}\cr
   &={H_o^4 {\cal F} (\Omega)\over c^4} 
      \int_0^{\infty} dk\, k^2\, P(k)\, 
      F_C(k,\theta),\cr}\eqn\wolfe
$$
where $y=2R_h\sin (\theta /2)$ and $R_h=2c/H_o \Omega$ (Sachs and
Wolfe, 1967). 
Strictly speaking, this is the correct expression only
for the Einstein-deSitter $\Omega=1,\ \Lambda=0$ universe, where
$\Lambda$ is the cosmological constant. When $\Lambda =0$ and
$\Omega\not= 1$, it
is easy to generalize this for small enough angles corresponding to
scales which crossed the horizon at early times when spatial curvature
had little effect on the evolution of the Universe. Hence, for $\Omega
< 1$, expression \wolfe\ is a good approximation for small angles
$\theta < \Omega $ radians. 
In most cases, the Sachs-Wolfe effect 
is the dominant contributor to the background
fluctuations for $\theta \gg 30' \Omega^{1 / 2}$; at these large
angles,  scattering
effects are negligible.

A slight variant of equation \wolfe\ also remains a good approximation
in the case of a zero curvature model with $\Lambda = 3(1-\Omega
)H_o^2/c^2$ (Kofman and Starobinsky 1985, Peebles 1984). Again, for small
angles, Peebles (1984) showed that $R_h$ should be replaced by the
comoving distance
$$
   R_\Lambda = \frac{c}{H_o\Omega^{1/2}} \int_1^\infty dy
   \left( y^3 +\Omega^{-1} -1 \right)^{-1/2} , \eq
$$
and that the expression ${\cal F} (\Omega)$ should be replaced by
an integral, which can be approximated as $\Omega^{1.54}$ (Efstathiou
\etal 1992). So that our
results can be valid for all these cases, we will investigate the
kernel $F\subr{C} (k,\theta )$, defined in equation \wolfe.

Expanding $F_C(k,\theta)$ in terms of Legendre polynomials shows the
magnitude of the multipole contributions.
$$
   F_C(k,\theta)=k^{-4} \sum_{l=0}^\infty (2l+1)\left[j_l(kR_h)\right]^2
     P_l(\cos\theta)= \sum_{l=0}^\infty f_l(k)  P_l(\cos\theta)
    , \eqn\fc
$$
where $j_l$ are spherical bessel functions and $P_l$ are Legendre polynomials. 
Figure 2 shows the amplitude of each of these contributions to the
kernel (for fixed
$\theta$) as a function of wavenumber. 
The curves are each renormalized to a maximum of
unity, and the
monopole and dipole terms, which are singular at $k=0$, are  not
shown. Clearly the 
dominant contribution to the integral in \wolfe\ for 
small $k$ (corresponding to large
spatial scales) is from the monopole and dipole moments as all others
are finite at the origin. As Figure 2 shows, each 
subsequent, low order moment has a specific wavenumber 
associated with it, and
picks out a small range of wavenumbers in the power spectrum $P(k)$.
In particular, the quadrupole contribution is dominated be the very
shortest wavenumbers.
The position of the maximum is a monotonically increasing
function of $l$, while the amplitude decreases with $l$.
The distance in $k$-space between the maxima of adjacent $l$ tends to a
constant value while the width of the peak increases. Hence the
strong association of a given wavenumber with a given term in the
expansion weakens significantly for higher order terms.

In a similar fashion to equation \fc, one can expand the spatial
kernel in terms of Legendre polynomials:
$$
     G\subr{C}(\theta,r)=\sum_{l=0}^\infty g_l(r) P_l(\theta). \eq
$$
One normally ignores the monopole ($l=0$) term since it represents
only a renormalization of the mean temperature. Each of the $g_l(r)$
is related to the corresponding $f_l(k)$ from equation \fc\ using the
integral \gfrel.
Defining $x=r/2R\subr{h}$, the dipole term is given by
$$
        g_1(r) =R_h
        \cases {{ 1 \over 10} -{ x^2\over 6}+{ x^3\over 8} -{
        x^5\over 60}, &if $x < 1$ \cr {1 \over 24 x}, &otherwise.}
	\eqn\gdipole
$$
This term is normally ignored as well in cosmological comparisons
since the major contributor to the observed dipole is thought to be
the Doppler shift coming from the peculiar motion of our own galaxy. 
Doppler terms were explicitly excluded in Sachs and Wolfe's original
calculation.
The fact that contributions to the dipole can come from scales larger
than the horizon lends credence to the argument (Paczynski and Piran,
1990) that it is possible for a significant part of the dipole moment 
not to be Doppler in origin. 

The higher order moments are all zero for $x>1$. For $x<1$ and $l>1$
we find the general result
$$\eqalign{
	&g_l(r)=R_h\Biggl[{1 \over 2(2l+3)(2l-1)}-{1 \over 6}x^2\cr
	&+{(2l+1) \over 12}\sum_{k=0}^{l}
	{(-1)^k(l+k)! \over (l-k)!(k!)^2} \left(
	{2 \over 2k+1}-{3 \over 1+k}+{6 \over 2k+3}-{1 \over k+2}
        \right)x^{2k+3}\Biggr].\cr}
	\eqn\gl
$$
Figure 3a shows plots of these 
spatial kernels. The curves are again renormalized to a maximum
of unity, and the monopole term, which is singular, is not plotted. 
All moments of higher order than the dipole are zero beyond the 
horizon and have $(l-1)$ zero points within it. 
As in the equivalent $k$-space plot (figure 2) the maximum value
decreases with increasing $l$. Here there is no specific
distance associated with any given $l$ as all the peaks are at the
origin. However, there is some suggestion to a characteristic {\it range}
in $r$ for each $l$ as the width of the peaks decreases
monotonically. 

Figure 3b plots the derivatives of the functions shown in figure 3a,
which are relevant for the $J_3$ integration.
Higher moments have maxima at progressively
smaller scales, just as implied by the case in $k$ space.
Essentially, higher moments arise from correlations (or mass
concentrations) at smaller spatial scales. Note that these curves have
once again been renormalized and the amplitude of the initial peak
actually decreases with $l$. 
Because correlations at a given angle $\theta$ involve sums over many
values of $l$, these plots (figures 3a and b) lead us to expect 
that such correlations do not arise from fluctuations on a specific associated
spatial scale, but rather from similar ranges in $r$.

One can also transform the general kernel, $F\subr{C} (\theta, k)$
defined in equation \fc. Defining $s\equiv \sin \theta /2$, 
the kernel for the correlation function with the monopole term removed
(to get rid of singularities) is
$$
        G_C(\theta,r)=R_h
        \cases {({1 \over 6}-{s \over 4}) + ({1 \over 6}
        -{1 \over 12s}) x^2 -{x^3 \over 24}, &if $0 < x < s$ \cr
        -{s^2 \over 12 x}+{1 \over 6} - {x \over 4} + {x^2 \over 6}
        -{x^3 \over 24}, &if $s < x < 1$ \cr
        (1-2s^2){1 \over 24x}, &  otherwise.} \eqn\gctr
$$
This agrees with a transform of the
function  derived from first principles by Traschen and Eardley (1986).
The derivative function is given by
$$
      \frac{\partial \ }{\partial r}  G_C(\theta,r)=
        \cases { ({1 \over 6}
        -{1 \over 12s}) x -{x^2 \over 16}, &if $0 < x < s$ \cr
        {s^2 \over 24 x^2} - {1 \over 8} + {x \over 6}
        -{x^2 \over 16}, &if $s < x < 1$ \cr
        -(1-2s^2){1 \over 48 x^2}, &  otherwise.} 
        \eqn\dgctr
$$

Contour plots of $G\subr{C} (\theta,r)$ and its derivative are shown
in Figures 4a and 4b respectively, with the dipole term also removed.
These figures demonstrate that there is very little tendency for
correlations at a given angle to correspond to a given spatial
distance $r$ for either the integral involving $\xi (r) $ or the one
involving $J_3(r)$. Such a tendency may exist for very small angles of
a few degrees or less, but for the most part, it is absent.

This is further demonstrated in Figures 5 and 6. In Figures 5a and 5b we show
slices through the contour maps in Figures 4a and 4b. Essentially all
of the curves (for all angles shown) give significant contributions to
$C(\theta )$ from similar regions of space. Figures 6a and 6b present
similar results but with the quadrupole term removed as well. Now the
curves have more structure, but there is again little or no tendency
for correlations on larger angles to arise from greater distances. 

\bigskip
\medskip
\line{{\bf 3~~ Conclusions.} \hfil}
\nobreak

We have derived explicit relations between the mass autocorrelation
function $\xi (r)$ and the angular correlation of background radiation
on the sky $C(\theta )$ due to the Sachs-Wolfe effect. We also derived
expressions relating  $C(\theta )$ (as well as its multipole moments)
to the second moment of $\xi (r)$, namely $J_3(r)$. We argued that the
integral involving $J_3 (r)$ is in some sense the more fundamental,
since unlike the autocorrelation function, $J_3(r)$ does not satisfy an
integral constraint. Because of the integral constraint, the kernel
$G\subr{C} (\theta ,r)$ does not provide the unique contribution from
every radius $r$ since every positive correlation must be balanced by
a negative correlation somewhere else.  $J_3(r)$ does not suffer this problem.

The expressions derived here may be used either with theoretical or
observed mass autocorrelation functions to derive the microwave background
fluctuations. Of course, it is true that large galaxy surveys can be
used to derive either an autocorrelation function or a power spectrum
(Davis and Peebles 1983, Maddox \etal, 1990). Hence either 
equation \wolfe\ or equation \gctr\ can in principle
be used to derive the anisotropy. However, the spatial integral
provides a direct measure of the volume that must be surveyed to
obtain an accurate result at a given angle. If the dipole term
(equation \gdipole) is also
removed from the kernel it does not require
integration over scales larger than the horizon.

The most surprising result that we find is  that mass fluctuations on all 
spatial scales contribute to temperature fluctuations on any but the
very smallest angular
scales, as is demonstrated in Figures 4 and 5. This arises in part
because the correlation at any angle is the sum of contributions from
many multipole moments. Although there is a clear tendency for the
higher multipole moments to arise from smaller distances $r$, when one
adds the many multipole moments to obtain $G\subr{C}$, this tendency
disappears. Hence it is dangerous and incorrect to assume that each angle
$\theta$ is related to a unique distance scale $r$ for any correlation.

\medskip

We would like to thank Luiz DaCosta and Sandra Faber for helpful
discussions. This work was supported in part by grants from NSF, NASA,
and a UCSC faculty research grant.

\vfill
\eject

\line{{\bf Figure Captions} \hfill}

\qq
{\bf Figure 1} {\bf a.} Derived kernels for mass fluctuations and bulk
velocities. $R^3\,G_{\delta M}$ plotted for a top-hat window function of
radius R (solid line) and a Gaussian window of the same volume
(dotted line). $R\,G\subr{v}/ H_o^2 {\cal F}^2(\Omega)$ plotted 
for a top-hat
window (short dashed line) and a Gaussian window (long dashed line).
{\bf b.} Kernels for integration over $J_3$ found by differentiating
the functions shown in Figure 1a.

\qq
{\bf Figure 2} Amplitude of multipoles of the $k$-space kernel $f_l(k)$ for
$l=2,3,..6$. These curves have been renormalized to a maximum of
unity. 

\qq
{\bf Figure 3} {\bf a.} Amplitude of multipoles for the real-space
kernel, $g_l(r)$ for
$l=1,2..6$. The curves have been renormalized to give a maximum of
unity. 
{\bf b.} Kernels for integration over $J_3$ found by differentiating
the functions shown in Figure 3a. The curves have been renormalized to 
give a maximum of unity. 

\qq 
{\bf Figure 4} {\bf a.} Contours of 
$G_C(r,\theta)/R_h^3$
with monopole and dipole only subtracted. The maximum occurs at
$\theta=r=0$ and has value ${1 \over 15}$. Negative contours are
indicated by dashed lines. Levels are shown at logarithmic intervals
with spacing of $.4 dex$ with the zero level contour in bold and the
lowest magnitude contour at $10^{-6}$.
{\bf b.} Derivatives of function shown in Figure 4a.


\qq
{\bf Figure 5} {\bf a.} The value of the kernel $G\subr{C} (\theta ,r
)/R_h^3$  (with the monopole and dipole removed)  for 
$s = \sin (\theta/2) = 0.1, 0.3, 0.5 , 0.7, 0.9$ as functions of r. These
are just slices through Figure 4a.
{\bf b.} Kernels for integration over $J_3$ found by differentiating
the functions shown in Figure 5a.

\qq
{\bf Figure 6} {\bf a.}  The value of the kernel $G\subr{C} (\theta ,r
)/R_h^3$ (with the monopole, dipole, and quadrupole removed) for
$s = \sin (\theta/2) = 0.1, 0.3, 0.5 , 0.7, 0.9$ as functions of r.
This is the same as Figure 5a but with the quadrupole
removed as well.  
{\bf b.} Kernels for integration over $J_3$ found by differentiating
the functions shown in Figure 6a.


\vfill
\eject

\line{{\bf References.} \hfill}

\qq
Blumenthal, G. R., da Costa, L. N., Goldwirth, D. S., Lecar, M.
and Piran, T. 1992 Ap. J.  388, 234.

\qq
Davis, M. and Peebles, P.J.E. 1983 Ap. J.  267, 465.

\qq
Dekel, A., Bertschinger, E., and Faber, S.M., 1990, Ap. J.  364,349.

\qq
Efstathiou, G., Bond, J. R. and White, S. D. M. M.N.R.A.S.  258, 1P.

\qq
Gorski, K. 1991), Ap. J. Lett.  370, L5.

\qq
Kashlinsky, A. 1992, Ap. J. 386, L37.

\qq
Kofman, L. A. and Starobinsky, A. A. 1985, Soviet Astr. Lett.
 11, 271.

\qq
Lightman, A.P. and Schechter, P.L., 1990, Ap. J. Suppl. 
74,831.

\qq
Lecar, M. 1993, Private Communication.

\qq
Lynden-Bell, D., Faber, S.M., Burstein, D., Davies, R.L., Dressler,
A., Terlevich, R.L. and Wegner, G., 1988, Ap. J.  326,19. 

\qq
Maddox, S.J., Efstathiou, G., Sutherland, W. J. and Loveday, J., 1990,
M.N.R.A.S.  242 43P.

\qq
Martinez-Gonzalez, E., and Sanz, J. L. 1989, Ap. J.  347,11.

\qq
Paczynski, B., and Piran, T., 1990, Ap. J.  364, 341.

\qq
Peebles, P. J. 1976, Ap. J.  205, 318.

\qq 
Peebles, P. J. E. 1981 Large Scale Structure in the Universe.
(Princeton University Press).

\qq
Peebles, P. J. E. 1984, Ap. J.   284, 439.

\qq
Sachs, R. K. and Wolfe, A. M. 1967, Ap. J.  147,73.

\qq
Smoot, G. F., Bennet, C. L., Kogut, A., Wright, E. L., Aymon, J.,
Boggess, N.W., Cheng, E.S., De Amici, G., Gulkis, S., Hauser, M.G.,
Hinshaw, G., Lineweaver, C., Loewenstein, K., Jackson, P. D., Janssen,
M., Kaita, E., Kelsall, T., Keegstra, P., Mather, J., Meyer, S. S.,
Moseley, S. H., Murdock, Y., Rokke, L., Silverbeg, R. F., Tenorio, L.,
Weiss, R. and Wilkinson, D. T.  1992, Ap.J.  388, 234.

\qq
Suto, Y., Gorski, K., Juszkiewicz, R. and Silk, J. 1988, Nature 
332, 328.

\qq	
Traschen, J. and Eardley, D. M. 1986, Phys. Rev. D  34, 6.

\qq
Wolfe, A.M., 1968, Ap.J.  156, 803.

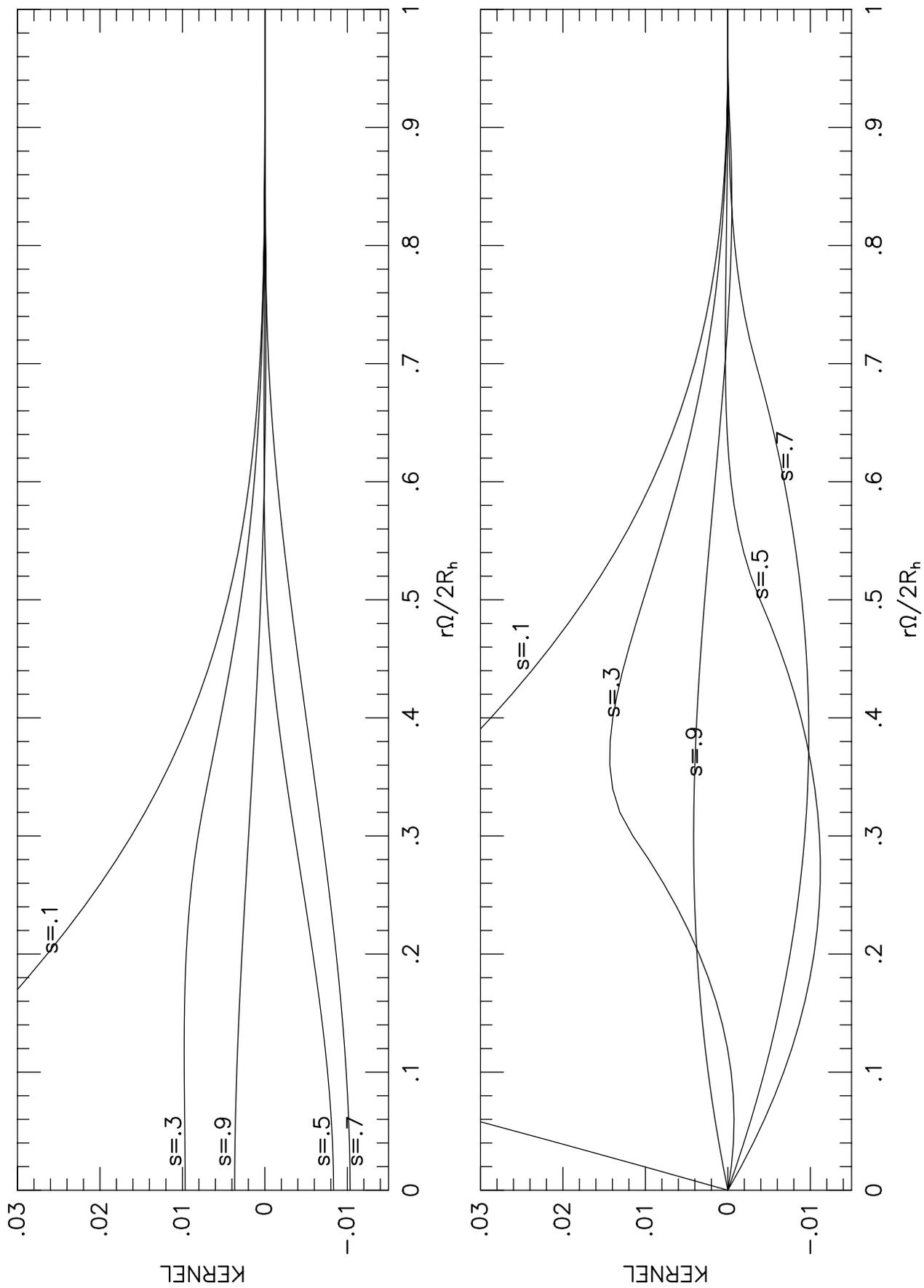
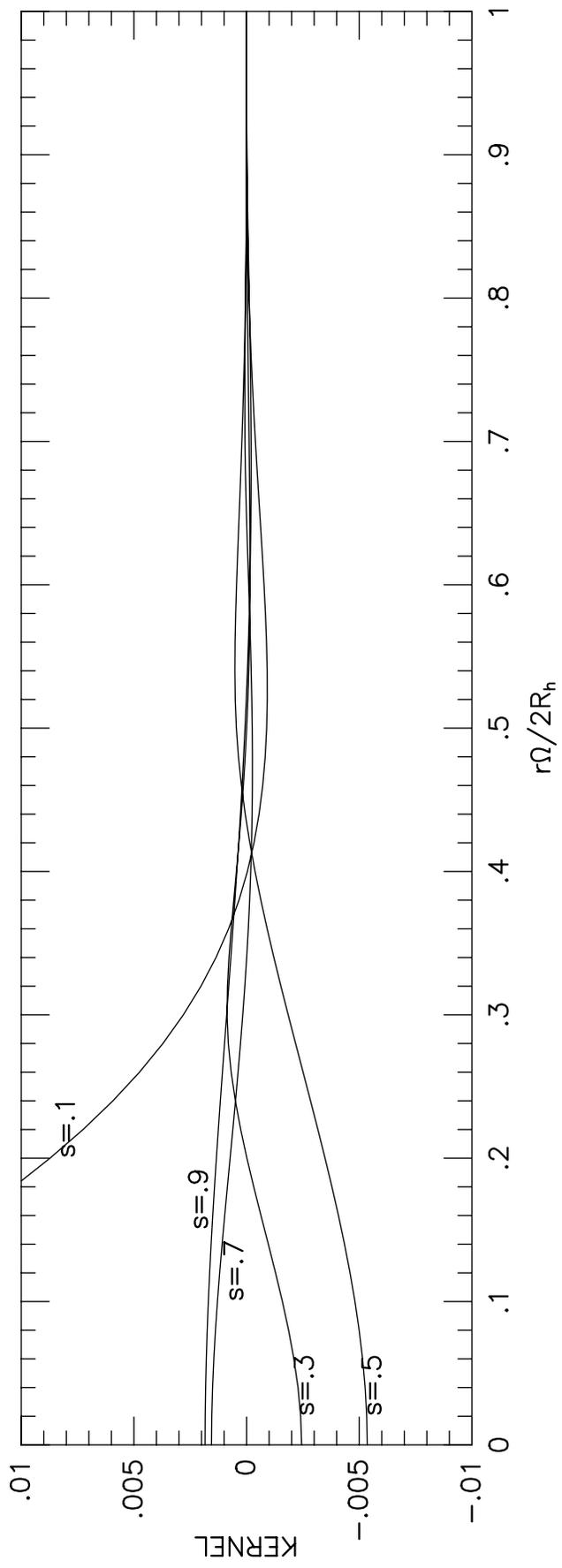
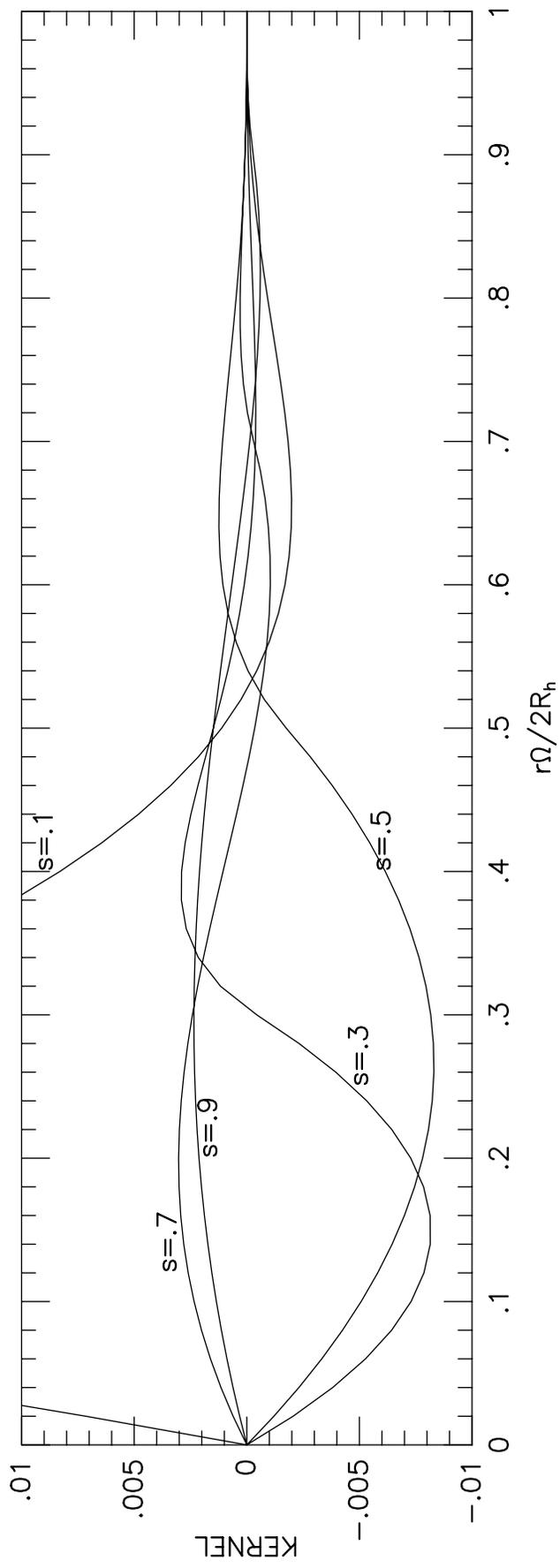
\bye